# Novel insights into the dynamics of intractable human epilepsy

Ivan Osorio*[§] M.D., Mark G. Frei[§] PhD., Didier Sornette[¶] PhD., John Milton[#], M.D., PhD.

* Department of Neurology, The University of Kansas Medical Center, 3599 Rainbow Boulevard, Mailstop 2012, Kansas City, KS 66160, USA. 913 (5884529)  email: iosorio@kumc.edu
§ 5040 Bob Billings Parkway, Suite A, Lawrence, KS  66049, USA. email: frei@fhs.lawrence.ks.us
¶ D-MTEC, D-PHYS and D-ERDW, ETH Zurich, Kreuzplatz 5, CH-8032 Zurich, Switzerland. email: dsornette@ethz.ch
# W.M. Keck Science Center, The Claremont Colleges 925 N. Mills Avenue, Claremont, CA 91711, USA. Email: jmilton@jsd.claremont.edu

**Summary**

Background: Knowledge of relevant statistical features such as probability distributions of energies (E) and inter-seizure intervals (ISI) of clinical and subclinical seizures is lacking in epileptology. This work endeavors to gain insight into the dynamics of Sz and their interactions by adopting a "systems" (non-reductionist) approach to their study, using powerful mathematical tools.

Methods: Probability density functions and the probability of Sz occurrence conditional upon the time elapsed from the previous Sz were estimated using the energy and intervals of SZ in prolonged recordings from subjects with localization- related pharmaco-resistant epilepsy, undergoing surgical evaluation.

Findings:  Clinical and subclinical seizure E and ISI distributions are governed by power laws in subjects on reduced doses of anti-seizure drugs.  There is increased probability of Sz occurrence 30 minutes before and after a seizure and the time to next seizure increases with the duration of the seizure-free interval since the last one. Also, over short time scales, "seizures may beget seizures".

Interpretation: The cumulative empirical evidence is compatible with and suggests that at least over short time scales, seizures have the inherent capacity of triggering other seizures.  This may explain the tendency of seizures to cluster and evolve into status epilepticus. Power law distributions of E and ISI indicate these features lack a typical size/duration and may not be accurate criteria or sufficient for classifying paroxysmal activity as ictal or interictal. This dependency and the existence of power law distributions raise the possibility that Sz occurrence and intensity may be predictable, without specifying the likelihood of success.

Key words: seizure energy; inter-seizure interval; intractable;  probability density function; conditional probability;  power law; seizure dependency; epilepsy dynamics

**Introduction**

The temporal behavior and other dynamical aspects of human epilepsy such as seizure duration and intensity (severity) is an underdeveloped area in epileptology. This is largely due to lack of complete and accurate data as seizure diaries, the current "gold standard", do not satisfy either of these conditions (refs). In the absence of dynamical knowledge of epilepsy at a large/macroscopic scale, characterization of its pathophysiology will remain incomplete. The seminal concept put forth by Hughlings Jackson in the $19^{th}$ century is that seizures are the result of an imbalance between excitation and inhibition, laid the foundations for the study of the cellular mechanisms of ictiogenesis that are likely to continue at ever smaller temporo-spatial scales for the foreseeable future. This approach has yielded valuable insights, but cannot in isolation, provide the knowledge necessary to further advance epileptology since it focuses only on the cardinal manifestation (seizures) while ignoring the disorder (epilepsy). Fundamental questions such as: "What are the probability distributions of times of seizure occurrences and of energies?" and, Are seizures independent of each other?" or, "Do seizures have the inherent capacity to trigger seizures?" remain unanswered, more than 100 years after Gowers made the clinical observation that "seizures beget seizures".  The dearth of knowledge of epilepsy dynamics may account in part for the fact that despite attempts spanning nearly two decades, worthwhile prediction of seizures remains elusive (1-7). Review of the various approaches to prediction reveals that all share in common a reductionist approach. That is, all attempts at forecasting the time of seizure occurrences have been based solely on contemporaneous changes in electrical signals recorded at or near the site of presumed ictiogenesis, thus ignoring potentially relevant temporal (the "history") and spatial (global/systems) information.

This work endeavors to gain insight into the dynamics (8) of localization-related pharmaco-resistant epilepsies by adopting a "systems" approach to its study and by using simple but powerful mathematical tools that have proven useful in fields that investigate the behavior of complex systems.

**Methods**

Three tools were used to investigate the dynamics of human intractable epilepsy: 1. Probability density or distribution functions (PDFs) of seizure energy (E) and of inter-seizure intervals, (ISI); 2. Superposition ("stacking") analysis and, 3. Empirical estimation of the probability of seizure occurrence conditioned upon the time elapsed from the previous seizure.

A probability density function (pdf) is a function from which the probability distribution for a random variable to take values in a given interval can be obtained by integration of the pdf over this interval. Loosely, a probability density function can be seen as a "smoothed out" version of a histogram. The PDF has an associated statistical measure that can provide constraints and guidelines to identify the underlying mechanisms of the behavior of complex systems such as the brain.

Conditional probability is the probability of some event A, given the occurrence of some other event B. Conditional probability is written P(A|B), and is read "the probability of A, given B".

Superposition analysis orderly "stacks" a variable using a "marker" to ensure alignment. In this analysis, seizures are the variables and their onset and termination times are the "markers" that allow precise their precise alignment.

With approval from the Human Subjects Committee and from each subject (signed consent form), quantitative analyses were performed on 16032 automated seizure detections (Sz) in prolonged (several days' duration) intracranial recordings from 60 human subjects with mesial temporal and frontal lobe pharmaco-resistant epilepsies on reduced doses of medications, undergoing evaluation for epilepsy surgery at the University of Kansas Medical Center (1996-2000). The vast majority of these seizures lacked behavioral manifestations and were classified as subclinical. Using a validated detection algorithm (9,10) Sz onset and end times, duration, intensity and site of origin were obtained. Szs were defined in humans as the dimensionless ratio of brain seizure to non-seizure electrical activity in a particular weighted frequency band reaching a threshold value, T, of at least 22 and remaining at or above it for at least 0.84 s (duration constraint, D) as previously described elsewhere. Two key variables were derived from these data: (1) "Energy" (E), defined as the product of each Sz's peak intensity ratio and its duration (in seconds) and (2) Inter-Sz event-interval (ISI), defined as the time (in seconds) elapsed between the onset of consecutive Szs. The reason for considering E and ISI lies in the fact that to optimize usefulness, Sz forecasts should include not only time of occurrence, but also their intensity, so that interventions, especially warning, may depend on whether the upcoming seizure is likely to be clinical or subclinical; warning for subclinical seizures may be optional so as to minimize anticipatory anxiety.

To characterize the statistical distribution of energy (E) and inter-seizure interval (ISI) of clinical and subclinical seizures, pooled values (all subjects) of E and ISI were used to construct doubly logarithmic plots. For this, the number of Szs of a given E and the number of ISI of a given duration D, were used to construct histograms whose bins were geometrically spaced (powers of 2) and made to span the entire range of the data. The number of seizures in each bin was then normalized by the bin's width and plotted on log-log scale.

Additionally, the temporal evolution of the probability of being in seizure as a function of time before the onset and following the termination of a given seizure, was investigated as follows: 1. The state of being in seizure was assigned a value of 1 and of being in the interictal state (non-seizure) a value of 0; 2. Using superimposed epoch analysis, the seizure onsets were "timelocked" to all other onsets and the ends of seizures to all other ends; 3. The state values, overlayed in this manner, were then averaged to compute the empirical probability, *P(t)*, of being in seizure at a relative time, *t*, in reference to the onset and termination of another seizure; 4. The resulting probability curves for each subject were then normalized by the subject's total fraction of time spent in seizure and averaged across all subjects.

## Results

Seizure Energy Distribution

The probability of a Sz having energy, E, larger than $x$ is proportional to $x^{-\beta}$, where $\beta \approx 2/3$ (Fig.1). This pdf differs from a Gaussian or normal pdf in its skeweness (to the right) that appears as a "heavy" or "fat" tail, reflecting the presence of very large ("extreme") events that occur with non-negligible probability. These "extreme" events lie many more "standard deviations" away from the "mean" than predicted by a Gaussian pdf. These properties are also reflected in the fact that unbounded power law distributions with $\beta \approx 2/3$ have infinite mean and variance.

Temporal Distribution of Seizures

In humans with pharmacoresistant epilepsies undergoing invasive monitoring, there is increased probability of Sz occurrence in the window beginning 30 minutes before a Sz and ending 30 minutes afterward (Fig. 2). That is, seizures in these subjects and under the conditions they were studied had a tendency to form clusters.

Distribution of inter-seizure intervals

The pdf estimates for inter-seizure intervals (ISI), defined as the time elapsed from the onset of one seizure to the onset of the next, were also calculated using histogram-based estimation methods. The pdf of ISI (Fig, 3) approximately follows a power-law distribution with $\beta \approx 0.5$. This distribution encompasses very short and long interseizure intervals, consistent with seizures clustering and prolonged seizure-free intervals in this population .

Paradox of conditional expected waiting time to next seizure

The prediction, derived from the heavy tail structure of the waiting time distributions between successive events (Fig. 3), which paradoxically implies that for such heavy-tailed distributions, "the longer it has been since the last event, the longer the expected time till the next', (11) was tested. The results confirmed that for seizures as for earthquakes (11), the dependence of the average conditional additional waiting time until the next event, denoted $<\tau|t>$, is directly proportional to the time t already elapsed since the last event (Fig. 4). For Szs, for short times t since the last event, $<\tau|t>$ is smaller than the average (unconditional) waiting time $<\tau>$ between two events, and then increases until it becomes significantly larger than $<\tau>$ as t increases. This means that the longer the time since the last seizure, the longer it is expected to be until the next one.

## Discussion

The temporal dynamics of seizures originating from discrete brain regions in subjects with pharmaco-resistant epilepsies on reduced doses of antiseziure medications may be partly described by laws, more specifically, by power laws that bear striking similarity to those governing seismic activity (12). The existence of a power law for E (the product of seizure intensity and duration) indicates that if E is the energy released during a seizure, the probability of occurrence of a seizure of intensity $x$ or larger is proportional to $x^{-\beta}$, being much larger for mild than for severe seizures; the same pdf and interpretation applies to ISIs. Power laws, which are ubiquitous in nature, are endowed with the property of "self-similarity" or "scale-invariance". This means that the shape of the distributions of physical quantities such as E and ISI does not change with changes in the scale of observation. Consequently, there is no typical Sz energy or ISI, but a "continuum" of E (sizes) and waiting times between events (ISI). The clinical implication of this scale invariance in this case, is that intensity or duration may not be fundamental or defining seizure properties. That is, and contrary to the universally sanctioned practice in basic and clinical epileptology, intensity and/or duration may not be accurate criteria to classify certain neuronal activity as either seizure or interictal (non-seizure). At a more abstract level, scale invariance in seizures may be conceptualized as the hallmark of certain complex systems (the brain in this case) in which, at or near a "critical" point/threshold (for ictiogenesis), its component elements are correlated over all existing spatial scales (neuron, minicolum, column, macrocolum, etc.) and temporal scales (microsecond, millisecond, second, etc). The characteristic of scale invariance is for example, also shared by cancer in which coupled mechanisms interact across multiple spatial and temporal scales: from the gene to the cell to the whole organism, from nanoseconds to years (13,14). Seizures' proclivity to entrain or "kindle" other brain regions may partly reflect the existence of multi-scale spatio-temporal correlations.

The value, $\beta \approx 2/3$, of the exponent of the power law of seizure energy, E, is indicative of a heavy-tail/extreme event distribution as opposed to a normal (Gaussian) distribution, and has interesting statistical and clinical implications: the mean and variance of distribution of E, are not definable, as their values are infinity for unbounded distributions in the ideal mathematical limit of infinite systems. In practice, for finite systems such as the brain, this means that the empirical determinations of the mean and of the variance of the distribution of E do not converge as they remain random variables sensitive to the specific realization of the data and in particular to the largest measured value. This is in stark contrast to the good convergence properties of the mean and variance in normally distributed random variables. This characteristic exponent, ($\beta \approx 2/3$), and more precisely the heavy tail, explains at a mathematical-conceptual level, the brain's tendency and capacity to support status epilepticus (SE), a form of extreme seizure. The other "path" to SE is through very short ISIs, which abound in the corresponding power law distribution (Fig. 3). Simply put, SE occurs when the brain's ictal activity "visits" the far right of the E distribution or the far-left region of the ISI power law distribution.

The structural and functional substrate to support the scale-free behavior observed in these seizure time series is in place: 1. The brain is an assembly of coupled, mainly nonlinear oscillators (neurons) with labile and unstable dynamics (15) and the length, density/clustering and patterns of neuronal interconnectivity have fractal or self-similar properties, that are repeated across a vast hierarchy of spatial scales (16-19); 2. Human magnetoencephalographic data obtained during rest and tasks revealed that large-scale functional neuronal networks that generate delta, theta, alpha, beta, and gamma frequency rhythms have attributes that are preserved across these frequency bands and that flexibly adapt to task demands. The most remarkable characteristic of these networks is their relative invariance of the network topology across all physiologically relevant frequency bands, forming a self-similar or fractal architecture (20). Dynamical analysis showed that these networks were located close to the threshold of order/disorder transition in all frequency bands and that behavioral state did not strongly influence global topology or synchronizability. Similarly, human EEGs showed scale-free dynamics and self-similar properties during eyes-closed and eyes-open, no-task conditions, but the scaling exponent differed significantly for different frequency bands and conditions (21). Those studies and this one suggest that scale-free behavior in the human brain may be insensitive to state (physiological vs. pathological such as in seizures), but its scaling factor may be sensitive to the prevailing conditions. For example, the slope of the pdf of size of spontaneous neuronal "avalanches" recorded 'in-vitro" changes from -3/2, to -2, in response to excitation with picrotoxin a $GABA_A$-receptor antagonist (22), but conserves its power law behavior. While power laws are often generated in systems with Euclidian geometry, the fractal geometry (16-19) of the brain likely illustrates a remarkable coupling between structure and processes, suggesting neural self-organization that generates both power laws statistics and dynamics as well as the fractal geometry of the brain. It is hypothesized that this fractal geometry and the emergent dynamics are intrinsically coupled as for seismogenic faults and earthquakes, implying that a successful prediction scheme for Sz requires understanding (as for earthquakes) of the interplay between dynamics at the time scale of sequences of Sz and of the structural elements of the brain.

The increased probability of pharmaco-resistant seizures to occur in clusters (Fig. 2), and the decreased probability of seizure occurrence with increasing time from the last one (Fig. 4), may be interpreted as: a) reflective of the inherent capacity of seizures to trigger seizures, thus supporting [at least over short time scales (minutes)] the concept put forth in the 19$^{th}$ century by Gowers (23) that "seizures beget seizures" and advanced by Morrell (24) in the second half of last century; b) indicative of some form of seizure interdependency or plasticity ("memory") in the system, as recently shown (25), and c) a harbinger of predictability, alluding to the possibility that seizures may be predictable, but without specifying the probability or ease of success. That seizures may be predictable is in itself a valuable finding for which no factual support had been sought, as those working in this field presumed predictability a priori. While at this juncture, seizure "predictability" cannot be generalized to out-of-hospital conditions and fully/properly medicated subjects, these findings, justify and foster, not only renewed efforts in the field of prediction, but also different approaches from those (6) applied to date. In particular, and at a minimum, the monitoring of observables should be expanded from the local (epileptogenic zone) to the global/systems scale and encompass both clinical and subclinical seizures, including their severity and the system's history. Prolonged ECoG recordings, from humans with pharmaco-resistant epilepsy contain frequent, low intensity short duration seizures that go unperceived by patients and observers ("subclinical") and have been consistently ignored for seizure forecasting purposes (25); these "subclinical" seizures should be included along with clinical seizures in prediction models (31).

Taken in their totality, these findings and the proposed systems ("non-reductionist") approach seem not only fruitful as evidenced by the uncovering of "laws" governing the temporal behavior of seizures and their energy distribution, but may also serve as the bases for expanding the inquiry into the dynamics of pharmaco-resistant epilepsy. This research direction may provide much needed impetus for the development of new or the refinement of existing theories and tools for the eventual control or prevention of epilepsy and mitigation of its negative psycho-social impact

Acknowledgements: This work was supported in part by NIH/NINDS grant nos. 5R21NS056022 and 1R01NS046602. JM is supported by the William R Kenan, Jr Foundation and NSF (grant no. 0617072).

None of the authors has a conflict of interest.

**REFERENCES**

1. Osorio I, Harrison MAF, Lai YC, Frei MG. Observations on the application of the correlation dimension and correlation integral to the prediction of seizures. J Clin Neurophysiol 2001; 18:269-74.

2. Lai Y-C, Osorio I, Harrison MAF, Frei MG. Correlation-dimension and autocorrelation fluctuations in seizure dynamics. Physical Review E 2002; 65:031921.

3. Lai Y-C, Harrison MAF, Frei MG, Osorio I. Inability of Lyapunov exponents to predict epileptic seizures. Phys Rev Lett 2003; 91:068102.

4. Harrison MAF, Frei MG, Osorio I. Accumulated energy revisited. Clin Neurophysiol 2005; 116:527-31.

5. Harrison MAF, Osorio I, Frei MG, Lai Y-C. Correlation dimension and integrals do not predict epileptic seizures. Chaos 2005; 15:33106.

6. Mormann F, Andrzejak RG, Elger CE, Lehnertz K. Seizure prediction: the long and winding road. Brain 2007; 130:314-33.

7. Lehnertz K, Mormann F, Osterhage et al. State-of-the-art of seizure prediction. J Clin Neurophysiol 2007; 24:147-53.

8. Milton J. Medically intractable epilepsy. In: Epilepsy as a dynamic disease (J. Milton, P. Jung, eds) Springer, New York, 2003, pp. 1-14.

9. Osorio I, Frei MG, Wilkinson SB. Real-time automated detection and quantitative analysis of seizures and short-term prediction of clinical onset. Epilepsia 1998; 39:615-27.

10. Osorio I, Frei MG, Sunderam S, et al. Performance re-assessment of a real-time seizure detection algorithm on long ECoG series. Epilepsia 2002; 43:1522-35.

11. Sornette D, Knopoff L. The paradox of the expected time until the next earthquake. Bull Seis Soc Am 1997; 87:789-98.

12. Osorio I, Frei MG, Sornette D, Milton J, Lai Y-C.: Seizures and earthquakes: Universality and scaling of critical "far from equilibrium" systems (Submitted to Phys Rev Letters)

13. Ribba B, Saut O, Colin T, Bresch D, Grenier E, Boissel JP. A multiscale mathematical model of avascular tumor growth to investigate the therapeutic benefit of anti-invasive agents. : J Theor Biol 2006; 243:532-41.

14. Schnell S, Grima R, Meini PK. Multiscale modeling in biology. American Scientist 2007; 95:134-45

15. Friston KJ. The labile brain. I. Neuronal transients and nonlinear coupling. Philos Trans R Soc Lond B Biol Sci 2000; 355:215-36.

16. Sporns O, Zwi JD. The small world of the cerebral cortex. Neuroinformatics. 2004; 2:145-62.

17. Breakspear M, Stam CJ. Dynamics of a neural system with a multiscale architecture. Philos Trans R Soc Lond B Biol Sci 2005; 360:1051-74.

18. Sporns O. Small-world connectivity, motif composition, and complexity of fractal neuronal connections. Biosystems. 2006; 85:55-64.

19. Honey CJ, Kötter R, Breakspear M, Sporns O. Network structure of cerebral cortex shapes functional connectivity on multiple time scales. Natl Acad Sci USA 2007; 104:10240-5.

20. Bassett DS, Meyer-Lindenberg A, Achard S, Duke T, Bullmore E. Adaptive reconfiguration of fractal small-world human brain functional networks. Proc Natl Acad Sci USA 2006; 103:19518-23

21. Stam CJ, deBruin DA. Scale-free dynamics of global functional connectivity in the human brain.. Hum Brain Mapp 2004; 22:97-109.

22. Beggs JM, Plenz D.  Neuronal Avalanches in Neocortical Circuits. J Neurosci 2003; 23:11167–77

23. Gowers, W.R.  Epilepsy and Other Chronic Convulsive Diseases: Their Causes, symptoms and Treatment. Churchill, London, 1901

24. Morrell F, deToledo-Morrell L. From mirror focus to secondary epileptogenesis in Man: AN Historical Review. In: Stefan H, Andermann F, Chauvel P, Simon S, eds. Advances in Neurology. Vol 81. Philadelphia: Lippincott Williams & Willkins, 1999:11-23

25. Sunderam S, Osorio I, Frei MG. Epileptic seizures are temporally interdependent under certain conditions. Epilepsy Res 2007; 76:77-84.

**Figure 1.** The probability density of pooled (all subjects) seizure energy E, (E = peak intensity x duration) approximately follows a power law with slope ≈ -2/3.   x-axis: seizure energy, E (logarithmic scale); y-axis: Number of seizures (logarithmic scale) with energy E.

**Figure 2.**  Empirical probability (0-1; y-axis) of being in seizure as a function of time elapsed (x-axis) before onset and after termination of a seizure. The curve to the left of the vertical dashed line (at time zero) depicts the probability before onset and the one to the right of this line, the probability after seizure termination. The empirical probability of being in seizure increases approximately 1200s before onset and returns to baseline 1200s after termination of a given seizure. This behavior is indicative of a strong clustering tendency.

**Figure 3.**  The probability density of pooled (all subjects) inter-seizure interval (ISI) approximately follows a power law with slope ≈ -1/2.   x-axis: Inter-seizure interval (logarithmic scale); y-axis: Number of inter-seizure intervals (logarithmic scale) with length L.

**Figure 4.** The expected time to next seizure (y-axis) increases  as a function of time elapsed since last (previous) seizure (x-axis). That is, the longer since last seizures the longer until the next seizure or viceversa.

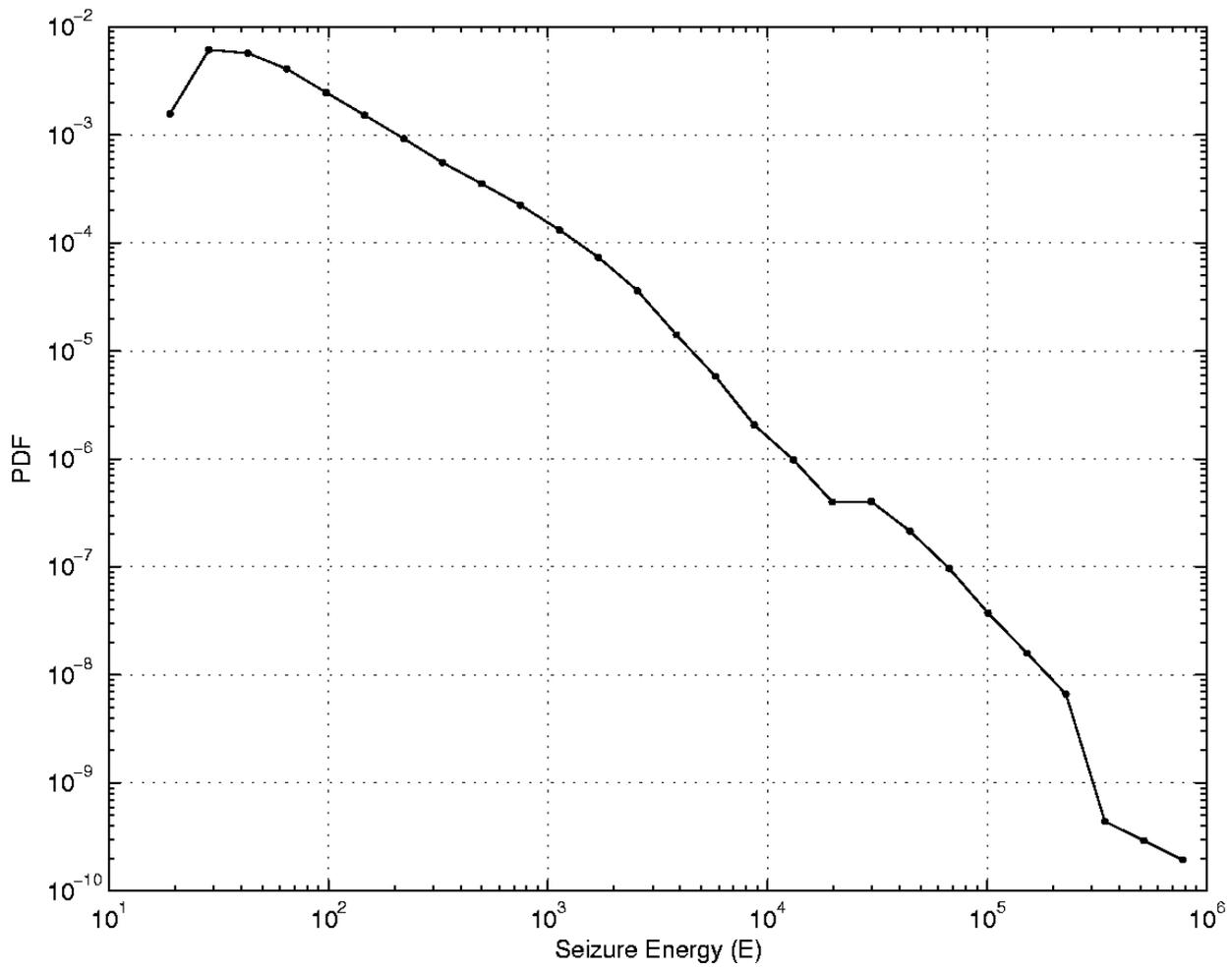

**Figure 1**

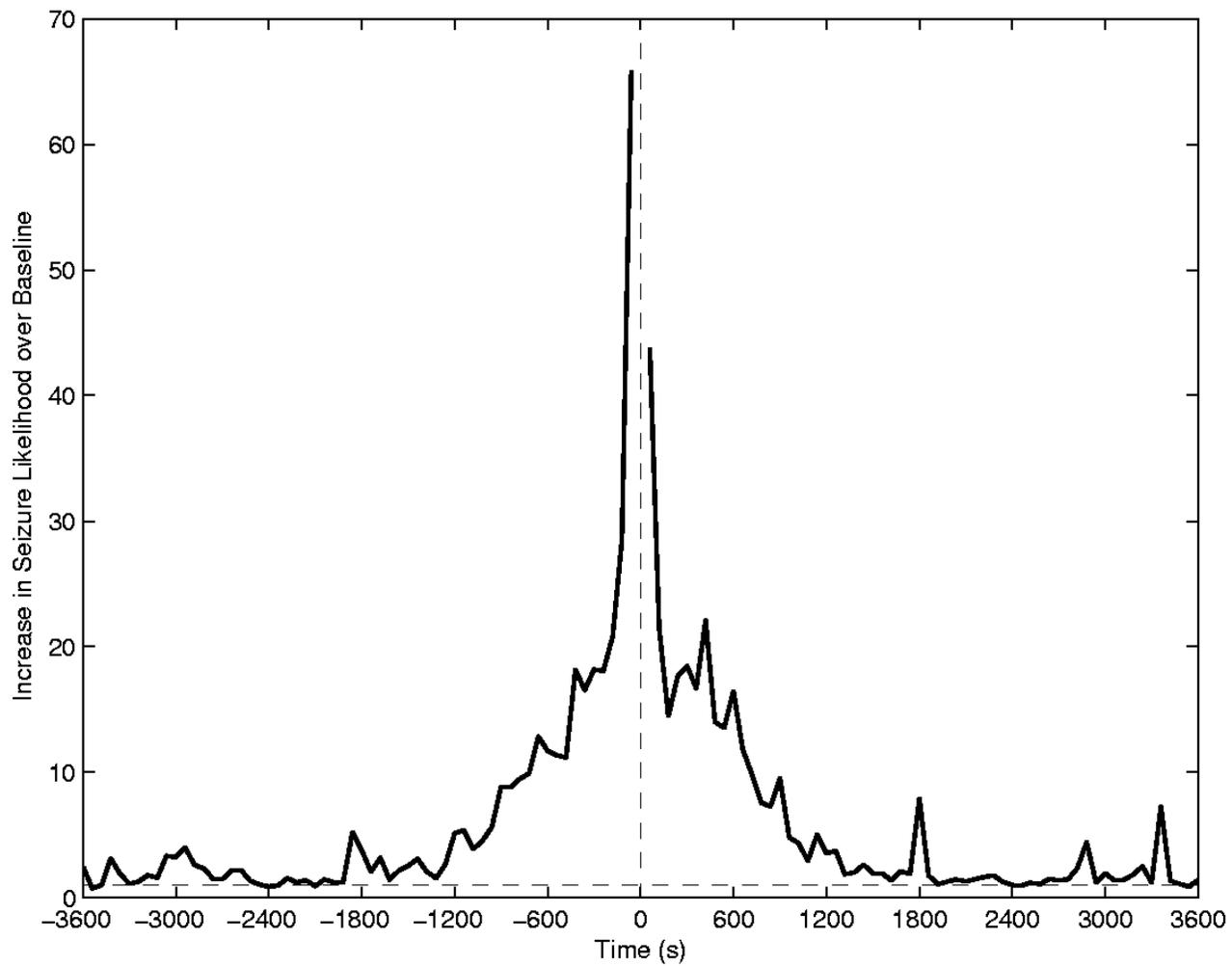

**Figure 2**

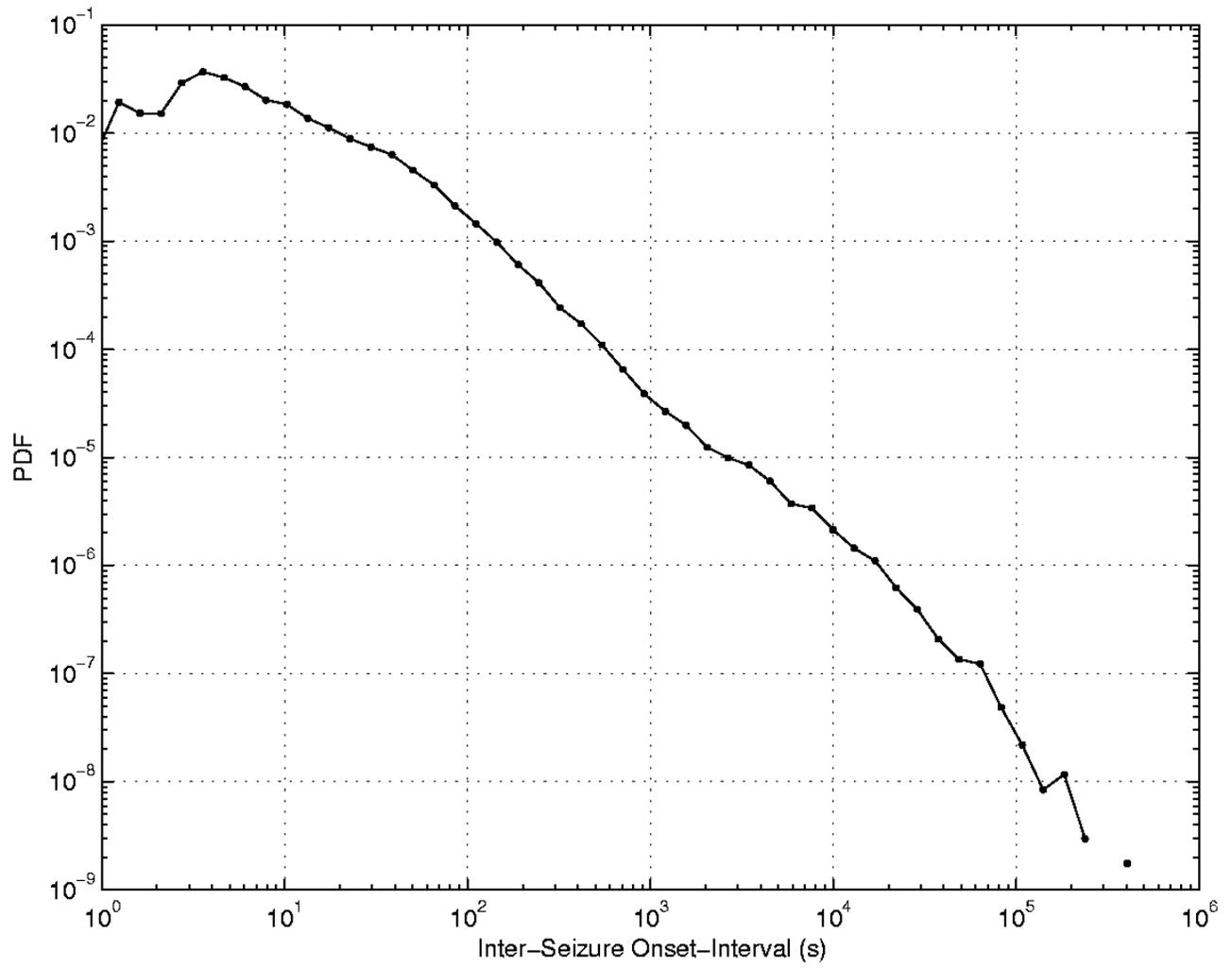

**Figure 3**

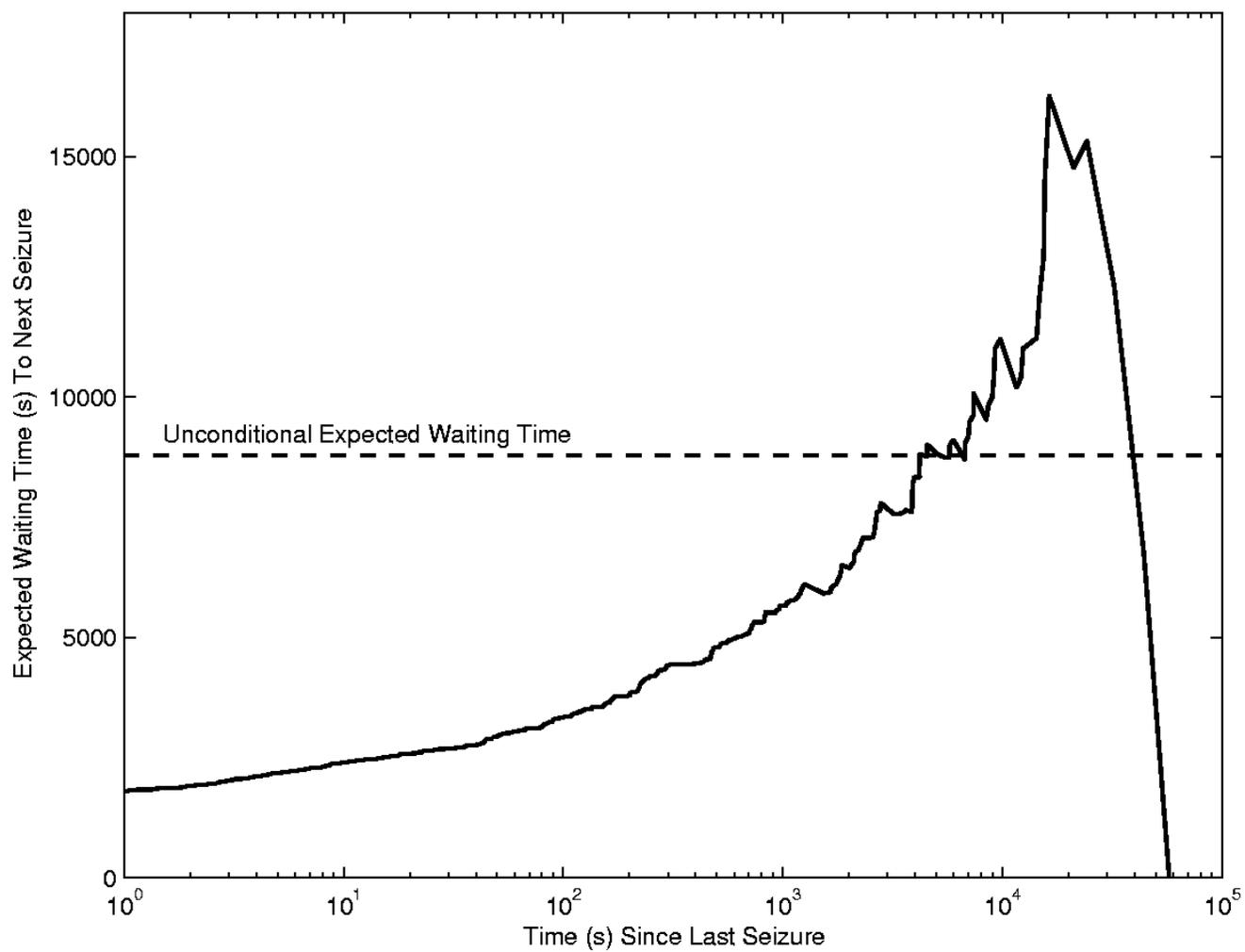

**Figure 4**